\documentstyle[11pt]{article}
\newcommand{\be}{\begin{equation}}
\newcommand{\ee}{\end{equation}}

\begin{document}
 
\noindent
{\large \bf  	THE PROBLEM OF CONSCIOUS OBSERVATION \hfill \break
IN QUANTUM MECHANICAL DESCRIPTION}\footnote{\ The first part of this
article is the slightly revised version of a paper with the same title
that was informally circulated in 1981 by means of the Epistemological
Letters of the Ferdinand-Gonseth Association in Biel (Switzerland) as
Letter No 63.0. (Therefore, terms such as ``new" or ``recent" in this
part refer to that year.)   In its Conclusion, this paper introduced
the term ``multi-consciousness interpretation" for a variant of the
Everett interpretation that has since
been rediscovered several times (more or less
independently), and become known as the ``many-minds
interpretation" of quantum theory.}
\vskip 1cm
\begin{quote}
\noindent
{\bf H. D. Zeh}
\vskip 0.2cm
\noindent
Institut f\"ur Theoretische Physik, Universit\"at Heidelberg,\\
Philosophenweg 19, D-69120 Heidelberg, Germany\\
\end{quote}
Epistemological consequences of quantum nonlocality (entanglement) are
discussed under the assumption of a universally valid Schr\"odinger
equation and the absence of hidden variables. This leads inevitably to
a {\it many-minds interpretation}. The recent foundation of
quasi-classical neuronal states in the brain (based on environmental
decoherence) permits in principle a formal description of the whole
chain of measurement interactions, including the {\it behavior} of
conscious observers, without introducing any intermediate classical
concepts (for macroscopic ``pointer states") or ``observables" (for
microscopic particle positions and the like) --- thus consistently
formalizing Einstein's {\it ganzer langer Weg} from the observed to
the observer in quantum mechanical terms.
\vskip .5cm

\noindent
Keywords: entanglement, quantum measurement, Everett interpretation
 
 
\section{Introduction}
 
John von Neumann seems to have first clearly
pointed out the conceptual difficulties that arise when one attempts to
formulate the physical process underlying subjective observation within
quantum theory \cite{1}. He emphasized the latter's
incompatibility with a psycho-physical parallelism, the traditional
way of reducing the act of observation to a physical process.
Based on the assumption of a physical
reality in space and time, one either assumes a ``coupling" (causal
relationship --- one-way or bidirectional) of matter and mind,
or disregards the whole problem by retreating to pure
behaviorism. However, even this may remain problematic
  when one attempts to describe classical
behavior in quantum mechanical terms. Neither
position can be upheld without fundamental modifications in
a consistent quantum mechanical description of the physical world.

These problems in formulating a process of observation within
quantum theory arise  as a consequence of quantum nonlocality
(quantum correlations or ``entanglement", characterizing generic
physical states), which in turn may be derived from the
superposition principle. This fundamental quantum property does
not even approximately allow the physical state of a local system
(such as the brain or parts thereof) to exist \cite{2}. Hence, no
state of the mind can exist ``parallel" to it (that is, correspond to
it one-to-one or determine it).

The question does not only concern the philosophical issue of
matter and mind. It has immediate bearing on quantum physics
itself, as the state vector seems to suffer the well known
reaction upon observation: its ``collapse". For this reason
Schr\"odinger once argued that the wave function might not
represent a physical object (not even in a statistical sense), but
should rather have a fundamental psycho-physical meaning.

This sitation appears so embarrassing to most physicists that
many tried hard to find a local reality behind the formalism of
quantum theory. For some time their effort was borne by the
hope that quantum correlations could be understood as
statistical correlations arising from an unknown ensemble
interpretation of quantum theory. (An ensemble explanation {\it
within} quantum theory can be excluded \cite{2}.) However, Bell's work
has demonstrated quite rigorously that any local reality --- regardless
of whether it can be experimentally confirmed in principle or not ---
would necessarily be in conflict with certain predictions of quantum
theory. Less rigorous though still quite convincing arguments had been
known before in the form of the dynamical completeness of the
Schr\"odinger equation for describing isolated microscopic systems,
in particular those containing quantum
correlations (such as many-electron atoms).

Although the evidence in favor of quantum theory (and against
local realism) now appears overwhelming, the
continuing search for a traditional solution may be
understandable in view of the otherwise arising epistemological
problems. On the other hand, in the absence of any empirical
hint how to revise quantum theory, it may be wise to accept the
description of physical reality in terms of non-local
state vectors, and to consider its severe consequences seriously.
Such an approach may be useful regardless of whether it will later
turn out to be of limited validity.

The conventional (``Copenhagen") pragmatic attitude of
switching between classical and quantum concepts by means
of {\em ad hoc} decisions does, of course, not represent a
consistent description. It should be distinguished from that
wave-particle duality which can be incorporated into the general
concept of a  state vector (namely, the occupation number
representation for wave modes). Unfortunately, personal tendencies for
local classical or for non-local quantum concepts to describe ``true
reality" seem to form the major source of misunderstandings
between physicists --- cf. the recent discussion between
d'Espagnat and Wei\ss kopf
\cite{4}.

It appears evident that conscious awareness must in some way
be coupled to local physical systems: our physical environment has to
interact with and thereby influence our brains in order to be
perceived. There is even convincing evidence supporting the
idea that {\em all} states of awareness reflect physico-chemical
processes in the brain. These neural processes are usually
described by means of classical (that is, local) concepts. One
may speculate about the details of this coupling on purely
theoretical grounds \cite{5}, or search for them experimentally
by performing neurological and psychological work. In fact,
after a few decades of exorcizing consciousness from
psychobiology by retreating to pure behaviorism, the demon now
seems to have been allowed to return \cite{6}. On closer
inspection, however, the concept of
consciousness {\em as used} turns out to be a purely behavioristic one:
certain aspects of behavior (such as language) are
rather conventionally associated with
consciousness. For
epistemological reasons it is indeed strictly impossible to derive the
concept of {\em subjective} consiousness (awareness)
from a physical world. Nonetheless, subjectivity need not
form an ``epistemological impasse" (Pribram's term \cite{7}),
but to grasp it may require combined efforts from
physics, psychology and epistemology.

\section{The Epistemology of Consciousness}

By inventing his malicious demon, Descartes demonstrated the
impossibility of {\em proving} the reality of the observed
(physical) world. This hypothetical demon, assumed to delude our
senses, may thereby be thought of as part of (another) reality ---
similar to an indirect proof.

On the other hand, Descartes' even more famous {\em
cogito ergo sum} is based on our conviction that the existence of
subjective sensations cannot be reasonably doubted. Instead of forming
an epistemological impasse, subjectivity should thus be regarded as an
epistemological gateway to reality.

Descartes' demon does not disprove a real physical world --- nor does
any other epistemological argument. Rather does it open up the
possibility of a {\em hypothetical} realism, for example in the
sense of Vaihinger's {\em heuristic fictions} \cite{8}. Aside from
having to be intrinsically consistent, this hypothetical reality has
to agree with observations (perceptions), and describe them in the most
economical manner. If, in a quantum world, the relation between
(ultimately subjective) observations and postulated reality should
turn out to differ from its classical form (as has often been
suggested for reasons of consistency), new non-trivial insights may
be obtained.

While according to
Descartes my own sensations are beyond doubt to me,  I cannot {\em
prove} other people's consciousness even from the presumption of their
{\em physical} reality. (This was the reason for eliminating it from
behavioristic psychology.) However, I may better (that is, more
economically) ``understand" or predict others' behavior (which I {\em
seem} to observe in reality) if I {\em assume} that they experience
similar sensations as I do. In this sense, consciousness (beyond
solipsism) is a heuristic concept precisely as reality. {\em There is
no better epistemological reason to exorcise from science the concept
of consciousness than that of physical reality.}

A consequence of this heuristic epistemological construction of
physical {\em and} psychic reality is, of course, that language
gives information about the speaker's consciousness.
This argument emphasizes the epistemologically derived (rather than
dynamically emerged) nature of this concept. However, only that part
of others' consciousness can be investigated, that manifests itself as
some form of behavior (such as language). For this reason it may
indeed be appropriate to avoid any fundamental concept of
consciousness in psychobiology. This requires that conscious behavior
(behavior as though being conscious) can be completely explained as
emerging --- certainly a meaningful conjecture. It would
have to include our private (subjectively experienced) consciousness
if a psycho-physical parallelism could be established. Only for such a
dynamically passive parallelism (or epi-phenomenalism) would the
physical world form a closed system that in principle allowed complete
reductionism.

Before the advent of quantum theory this ivory tower position of
physics could be upheld without posing problems. If, on the other hand,
the nonlocal quantum concepts describe {\em real} aspects of the
physical world (that is, if they are truly heuristic concepts), the
parallelism has to be modified in some way. Such a modification could
some day even turn out to be important for experimental psychobiology,
but it is irrelevant whenever nonlocality can be neglected, as for
present-day computers or most neural processes. However, the
quasi-classical activities of neurons may be almost as far from
consciousness as an image on the retina. The concept of ``wholeness"
--- often emphasized as being important for complex systems such as
the brain --- is usually insufficiently understood: in quantum theory
it is neither a mere dynamical wholeness (that is, an efficient
interaction between all parts) nor is it restricted to the system
itself. Dynamical arguments require a {\em kinematical
wholeness of the entire universe} (when regarded as composed of {\em
spatial} parts)
\cite{2}. It may be neglected for certain (``classical") aspects
only --- not for a complete microscopic description
that {\it may} be relevant for subjective
perceptions.

\section{Observing in a Quantum World}

One possible consequence of these problems that inevitably arise in
quantum theory would be to abandon the heuristic and generally
applicable concept of a physical reality --- explicitly \cite{9} or
tacitly. This suggestion includes the usual restriction to formal rules
when calculating probability distributions of {\em presumed}
classical variables in situations which are intuitively understood
as ``measurements" (but insufficiently or even inconsistently
distinguished from normal ``dynamical" interactions). Clearly, no
general description of physical processes underlying awareness could be
given in the absence of a physical reality, even though macroscopic
behavior (including the dynamics of neural systems) can be described
by means of the usual pragmatic scheme. This is quite unsatisfactory,
since subjective awareness has most elementary meaning without {\em
external} observation (that would be required in the
Copenhagen interpretation). Epistemologically, any concept of
observation must ultimately be based on an observing subject.

This ``non-concept"
of abandoning microscopic reality is not at all required, as has
been pointed out before \cite{2,10}. Instead, one may regard the
state vector as ``actual" and representing reality, since it
{\em acts} dynamically (often as a whole) on what is observed.
Moreover, in view of Bell's analysis of the consequences of quantum
nonlocality, it appears questionable whether anything, and what, might
be gained from inventing novel fundamental concepts (hidden variables)
without any empirical support. Two different solutions of the
measurement problem then appear conceivable: von Neumann's collapse or
Everett's relative state interpretation
\cite{11}. In both cases a (suitably modified)
psycho-physical parallelism can be re-established.

A dynamical collapse of the wave function would require nonlinear and
nonunitary terms in the Schr\"odinger equation \cite{12}.
They may be extremely small, and thus become effective only through
practically irreversible amplification processes occurring
during measurement-like events. The superposition principle would
then be valid only in a linearized version of the theory. While this
suggestion may in principle explain quantum measurements, it would not
be able to describe definite states of concsiousness unless the
parallelism were restricted to quasi-classical variables in the
brain. Since nonlinear terms in the Schr\"odinger equation lead to
observable deviations from conventional quantum theory, they should
at present be disregarded for similar reasons as hidden variables.
Any proposed violation of the superposition principle must be
viewed with great suspicion because of the latter's great and general
success. For example, even superpositions of different vacua have
proven heuristic (that is, to possess predictive power) in
quantum field theory.

The problems thus arising
when physical states representing consciousness are described within
wave mechanics by means of nonlinear dynamical terms could possibly be
avoided if these nonlinearities were themselves {\em caused} by
consciousness. This has in fact been suggested as a way to incorporate
a genuine concept of free will into the theory \cite{13}, but
would be in conflict with the hypothesis of a closed physical
description of the world.

If the Schr\"odinger equation is instead assumed to be universal and
exact, superpositions of states of the brain representing different
contents of consciousness are as unavoidable as Schr\"odinger's
superposition of a dead and alive cat. However, because of unavoidable
interaction with the environment, each component must then be
quantum correlated with a different (almost orthogonal) state of the
rest of the universe. This consequence, together with the way how we
perceive the world, leads obviously to a ``many-worlds" interpretation
of the wave function.\footnote{\ Everett
\cite{11} suggested ``branching" wave functions in order to discuss
cosmology in strictly quantum mechanical terms (without an external
observer or a collapse). I was later led to similar conclusions as a
consequence of unavoidable quantum entanglement
\cite{2} --- initially knowing neither of Everett's nor of Bell's
work.}  Unfortunately this name is misleading. The quantum world
(described by a wave function) would correspond to {\em one}
superposition of myriads of components representing {\em classically}
different worlds. They are all dynamically coupled (hence
``actual"), and they may in principle (re)combine as well as branch. It
is not the real world (described by a wave funtion) that branches in
this picture, but consciousness (or rather the state of its physical
carrier), and with it the {\it observed} (apparent) ``world"
\cite{2}. Once we have accepted the formal part of quantum theory,
only our experience teaches us that consciousness is physically
determined by  (factor) wave functions in certain {\em components} of
the total wave function.\footnote{\ It would always be possible to
introduce additional (entirely arbitrary and unobservable) variables
as a hypothetical link between the wave function and consciousness.
Given their (hypothetical) dynamics, the required quantum
probabilities can then be postulated by means of appropriate initial
conditions. An example are the classical variables in Bohm's pilot
wave theory
\cite{14}.} The existence of ``other" components (with
their separate conscious versions of ourselves) is a heuristic fiction,
based on the assumption of a general validity of dynamical laws that
have always been confirmed when tested. When applied to classical
laws and concepts, an analogous assumptions leads to the
conventional model of reality in space and time as an extrapolation of
what is observed. In the quantum model, {\em a collapse would represent
a new kind of solipsism}, since it denies the existence of these
otherwise arising consequences.

Everett related his branching to the practically
irreversible dynamical decoupling of components
that occurs when microscopic properties become correlated to
macroscopic ones. This irreversibility requires specific initial
conditions for the global state vector \cite{5}. Such initial
conditions will then, for example, also cause a sugar molecule
to permanently send {\em retarded} ``information" (by scattering
photons and molecules) about its handedness into the universe. In this
way, their relative phases become nonlocal, and thus cannot affect the
physical states of local conscious observers (or states of their
brains) any more. The separation of these components is dynamically
``robust". There is no precise localization of the
branch cut (while  a
genuine dynamical collapse would have to be {\em specified} as a
dynamical law).

Nonetheless, Everett's branching in terms of quasi-classical properties
does {\em not} appear sufficient to formulate a psycho-physical
parallelism. Neither would this branching produce a definite factor
state for some relevant part of the brain, nor does every
decoherence process somewhere in the universe describe conscious
observation. Even {\em within} a robust branch, most parts of
the brain will remain strongly quantum correlated with one another and
with their environment.

Everett's branchings represent objective measurements --- not
necessarily conscious observations. A parallelism seems to be based on
a far more fine-grained branching (from a local point of view) than
that describing measurements, since it should correspond one-one to
subjective awareness. The conjecture here is: does the (not
necessarily robust) branching that is conceptually required for
defining the parallelism then readily {\em justify} Everett's
(apparently objective) branching into quasi-classical worlds?

The branching of the global state vector $\Psi$ with
respect to  {\em two} different conscious observers ($A$ and $B$, say)
may be written in their Schmidt-canonical forms
\cite{5},
$$
\Psi = \sum_{n_A} c_{n_A}^A  \chi_{n_A}^A \phi_{n_A}^A
=   \sum_{n_B} c_{n_B}^B  \chi_{n_B}^B \phi_{n_B}^B \quad ,  \eqno(1)
$$
where $\chi^{A,B}$ are states of the respective physical carriers of
consciousness (presumably small but not necessarily local parts of the
central nervous system), while $\phi^{A,B}$ are states of
the respective ``rests of the universe". In order to
describe the macroscopic {\em behavior} of (human) observers, one has
to consider the analogous representation with respect to the states
$\tilde \chi$ of their whole bodies (or relevant parts thereof),
$$
\Psi = \sum_{k_A} \tilde c_{k_A}^A  \tilde \chi_{k_A}^A \tilde
\phi_{k_A}^A  =   \sum_{k_B} \tilde c_{k_B}^B  \tilde \chi_{k_B}^B
\tilde \phi_{k_B}^B
\quad .  \eqno(2)
$$
In particular, the central nervous system may be assumed to possess
(usually unconscious) ``memory states" (labelled by $m_A$ and $m_B$,
say) which are similarly robust under decoherence as the handedness of
a sugar molecule. Time-directed {\em quantum causality} (based on the
initial condition for the global wave function) will then force the
Schmidt states
$\tilde \chi^{A}$ and $\tilde \chi^B$ to approximately factorize in
terms of these memory states \cite{15},
$$
\Psi \Rightarrow \sum_{m_A\mu_A} \tilde c_{m_A\mu_A}^A  \tilde
\chi_{m_A\mu_A}^A
\tilde
\phi_{m_A\mu_A}^A  \approx   \sum_{m_B\mu_B} \tilde c_{m_B\mu_B}^B
\tilde
\chi_{m_B\mu_B}^B
\tilde \phi_{m_B\mu_B}^B
\quad ,  \eqno(3)
$$
where $\mu_A$ and $\mu_B$ are additional quantum numbers. The ``rest
of the universe" thus serves as a sink for phase
relations.

In general, the robust quantum numbers $m_A$ and $m_B$ will be partly
correlated --- either because of special interactions between the two
oberservers (communication), or since they have arisen from the
same cause (that is, from observations of the same event). These
correlations define the concept of objectivization in quantum
mechanical terms.

The genuine carriers of consciousness (described by the states $\chi$
in (1)) must {\em not} in general be expected to represent
memory states, as there do not seem to be permanent contents of
consciousness. However, since they may be assumed to interact
directly with the rest of the $\tilde
\chi$-system only, and since phase relations between different quantum
numbers
$m_A$ or
$m_B$ would immediately become nonlocal, memory
appears ``classical" to the conscious observer. Each robust branch in
(2), hence also each $m$-value, describes essentially an independent
partial sum of type (1) when observed \cite{16}. The emprirically
relevant probability interpretation in terms of quasi-classical
branches (including pointer positions) may, therefore, be derived from
a similar (but fundamental) one for the {\em subjective branching}
(with respect to each observer) that according to this interpretation
defines the novel psycho-physical parallelism.

As mentioned before, macroscopic {\em behavior} (including behavior as
though being conscious) could also be described by means of the
pragmatic (probalistic) rules of quantum theory. An exact Schr\"odinger
equation does not imply deterministic behavior of conscious beings,
since one  has to expect that macroscopic stimuli have
microscopic effects in the brain before they cause
macroscopic behavior. Thereby, interaction with the environment will
intervene. Everett's ``relative state" decomposition (1) with respect
to the subjective observer state
$\chi$ may then considerably differ from the objectivized branching
(3), that would be meaningful with respect to all conceivable
``external" observations. This description may even put definite
meaning into Bohr's vague concept of {\em complementarity}.

\section{Conclusion}

The multi-universe interpretation of quantum theory (which should
rather be called a {\em multi-consciousness interpretation}) seems to
be the only interpretation of a universal quantum theory (with an exact
Schr\"odinger equation) that is compatible with the way the world is
perceived. However, because of quantum nonlocality it requires an
appropriate modification of the traditional epistemological postulate
of a psycho-physical parallelism.

In this interpretation, the physical world is completely described by
Everett's wave function that evolves deterministically (Laplacean).
This global quantum state then defines an indeterministic (hence
``branching") succession of states for all observers. Therefore,
  the world itself {\em appears} indeterministic --- subjective in
principle, but largely objectivized through quantum correlations
(entanglement).

This quite general scheme to describe the empirical world is
conceptually consistent (even though the parallelism remains vaguely
defined), while it is based on the presently best founded physical
concepts. The latter may some day turn out to be insufficient, but it
is hard to see how any future theory that contains quantum theory in
some approximation may avoid similar epistemological problems. These
problems arise from the contrast between quantum nonlocality
(confirmed by Bell's analysis as part of {\em reality}) and the
locality of consciousness ``somewhere in the brain". Quantum concepts
should be better founded than classical
ones for approaching these problems.
 
\section{Addendum}

The above-presented paper of 1981 has here been rewritten
(with minor changes, mainly regarding formulations), since the solution
of the quantum mechanical measurement problem proposed therein has
recently gained interest, while the Epistemological Letters (which
were used as an informal discussion forum between physicists and
philosophers interested in the ``new quantum debate") are now
hard to access. The dynamical dislocalization of phase relations (based on
\cite{2,15})
referred to in this article in order to justify robust Everett
branches has since become better known as {\em decoherence} (see
\cite{giulini}), while the ``multi-consciousness interpretation"
mentioned in the Conclusion has been rediscovered on several
occasions. It is now usually discussed as a ``many-minds
interpretation"
\cite{albert,lockwood,donald,stapp,chalmers,whitaker}, but has also
been called a ``many-views"
\cite{squires} or ``many-perceptions" interpretation
\cite{page}.

The conjectured quasi-classical nature of those dynamical states of
neurons in the brain which may carry memory or can be investigated
``from outside" has recently been confirmed by quantitative estimates
of their decoherence in an important paper by Tegmark
\cite{tegmark}. To most of these states, however, the true physical
carrier of consciousness somewhere in the brain may still represent an
{\em external} observer system, with whom they have to interact in
order to be perceived. Regardless of whether the ultimate
observer systems are quasi-classical or possess essential quantum
aspects, consciousness can only be related to factor states
(of systems assumed to be localized in the brain) that appear in {\em
branches} (robust components) of the global wave function --- provided
the Schr\"odinger equation is exact. Environmental decoherence
represents {\em entanglement} (but not any ``distortion" --- of the
brain, in this case), while {\em ensembles} of wave functions,
representing various {\em potential} (unpredictable) outcomes, would
require a dynamical collapse (that has never been
observed).\footnote{\ A collapse would be {\em conceivable} as well in
Bohm's theory, where memory and objective thought seems to be still
described by neuronal {\em quantum} states, rather than by the
classical configurations which according to Bell
\cite{14} would have to describe physical states corresponding to
consciousness.}

An essential role of the conscious observer for the occurrence of
fundamental (though objective) quantum events was apparently suggested
already by Heisenberg in his early ``idealistic" interpretation of
{\em a particle trajectory coming into being by our act of observing
it}. Bohr, in his later Copenhagen interpretation, insisted instead
that classical outcomes arise in the apparatus during
irreversible {\em measurements}, which
he assumed not to be dynamically analyzable in terms of a microscopic
reality (cf. Beller \cite{beller}). This first {\em classical} link in
the chain of interactions that forms the observation of a quantum
system can now be identified with the first occurrence of decoherence
(globally described as a unitary but practically irreversible {\em
dynamical} process --- see
\cite{zeh99}).

However, Bohr's restriction of the applicability of quantum concepts
as well as Heisenberg's uncertainty relations were meant to establish
{\em bounds to a rational description of Nature}. (The popular
simplistic view of quantum theory as merely describing stochastic
dynamics for an otherwise classical world leads to the well known
wealth of ``paradoxes", which rule out any local description but have
all been {\em derived} from the superposition principle, that is,
ultimately from an entangled global wave function.) Von Neumann's
``orthodox" interpretation, on the other hand, is somewhat obscured by
his use of observables, which should have no {\em fundamental} place
in a theory of interacting wave functions.  His postulate of a
dynamical collapse representing conscious observations was later
elaborated upon by London and Bauer
\cite{london}, while Wigner
\cite{13} suggested an active influence of the
mind on the physical state (that should better not
affect objectively measurable probabilities). Stapp
\cite{stapp} expressed varying views on this problem, while Penrose
\cite{penrose} speculated that human thinking, in contrast to
classical computers, requires genuine quantum aspects (including
superpositions of neuronal states {\em and} the collapse of the wave
function).\footnote{\ There seems to be a certain confusion between
logical {\em statements} (that is, tautologies), which have no implicit
relation to the concept of time, and algorithmic {\em
procedures}, performed in time in order to {\em prove}
them. (Undecidable {\em formal} statements are meaningless, and hence
not applicable.) A dynamical collapse of the wave
function must not be regarded as representing ``quantum logic" (or
``logic of time"). This misconception appears reminiscent of the
popular confusion of {\em cause} and {\em
reason}.}

The Everett interpretation leads to its ``extravagant" (unfamiliar and
unobservable) {\em consequences}, because it does not invent any
new laws, variables or irrational elements for the sole purpose of
avoiding them.  Lockwood
\cite{lockwood} is quite correct when he points out the essential role
of decoherence for the many-minds interpretation (see also
\cite{whitaker}). This unavoidable ``continuous measurement" of all
macroscopic systems by their environments (inducing
entanglement) was indeed initially discussed
\cite{2} precisely in order to support the concept of a universal wave
function, in which ``branching components" must be {\em separately}
experienced.

Heisenberg once recalled \cite{heisenberg} that Einstein had told him
(my translation): ``Only the theory may tell us what we can
observe.
\dots \ On the whole long path ({\it ganzer langer Weg}) from the event
to its registration in our consciousness you have to know how Nature
works." Einstein did thus {\em not} suggest that the theory has to
postulate ``observables" for this purpose (as the first part of this
quotation is often understood). Formal observables are useful only
since the subsequent part of this chain of interactions can {\em for
all practical purposes} be described in terms of classical variables,
after initial values have
been stochastically {\em created} for them somewhere in the chain.
However, most physicists would now agree on what to do (in principle)
if quantum effects should be relevant during some or all
intermediary steps (cf.
\cite{ghirardi}): they would have to calculate the evolution of the
corresponding series of entangled quantum systems, taking into account
decoherence by the environment where required. There is then no need
for genuine classical variables anywhere, since Tegmark's decohered
neuronal (quantum) states form an appropriate ``pointer basis" for the
application of quantum probabilities. These
probabilities need not characterize a stochastic dynamical process (a
collapse of the wave function), but would describe an objectivizable
splitting of the (state of the) mind if the Schr\"odinger equation
were exact. In Bohm's quantum theory, on the other hand, states of the
mind would be related to ``surreal" classical trajectories which are
guided by --- hence would be {\em aware} of --- a branch wave function
only
\cite{14}.

Therefore, I feel that the Heisenberg-Bohr picture of quantum
mechanics can now be claimed dead. Neither classical concepts, nor any
uncertainty relations, complementarity, observables, quantum logic,
quantum statistics, or quantum jumps have to be introduced on a
fundamental level (see also Sect.~4.6 of \cite{zeh99}). In a recent
experiment \cite{zeilinger}, interference experiments were
performed with mesoscopic molecules, and proposed for small
virus. Time may be ripe to discuss the consequences of similar
{\em Gedanken} experiments with objects carrying some primitive form of
``core consciousness" \cite{damasio} --- including an elementary
awareness of their path through the slits. How can ``many minds"
be avoided if their coherence can be restored?

I wish to thank Erich Joos for various helpful comments.
 
 

\end{document}